# The local structure of electrically stressed liquid water and implications for modelling of dielectric relaxation

Adam D. Wexler[1*], Brigitte Bitschnau[2], Antonio Cervellino[3], Nicola Casati[3], Alan K. Soper[4], Jakob Woisetschläger[5], and Elmar C. Fuchs[1]

[1] Wetsus, European Center of Excellence for Sustainable Water Technology, Oostergoweg 9, 8911MA, Leeuwarden, Netherlands.

[2] Institute of Physical and Theoretical Chemistry, Graz University of Technology, Rechbauerstraße 12, 8010 Graz, Austria

[3] Swiss Light Source, Paul Scherrer Institute, 5232 Viligen PSI, Switzerland

[4] ISIS Facility, STFC Rutherford Appleton Laboratory, Harwell Science and Innovation Campus, Didcot, Oxon, OX11 0QX, United Kingdom

[5] Institute for Thermal Turbomachinery and Machine Dynamics, Working Group Metrology – Laser Optical Metrology, Technical University of Graz, Inffeldgasse 25/A, 8010 Graz, Austria

* Corresponding author. E-mail: adam.wexler@wetsus.nl, phone: +31 (0) 58 2843 013

**Abstract**

In a floating water bridge the total radiation scattering of water stressed by a moderately strong electric field (1mV/nm) was compared to water without an applied electric field using X-ray and small angle neutron scattering. Structure refinement was carried out using the EPSR method and the TIP4P/2005 water model. These results did not reveal a significant difference in the local static structure of water however analysis of the simulation indicated that the modeled local potential energy surface reveals a departure between electrically stressed and unstressed water. The observed differences show that the local environment is changed by the applied electric field although weak relative to the intermolecular coulombic field. When discussing the results we show that the current methods used to simulate the pair potentials are still insufficient to treat such non-equilibrium systems and further simulation techniques have to be developed to properly reconstruct the microscopic dielectric relaxation process.

**Introduction**

Understanding structure in liquids where dynamics have equal or greater influence on the continuum level properties is a formidable and ongoing task [1]. Unlike solid state systems where the location of the atoms remains relatively fixed, in liquids the rapid and relentless molecular motion limits the definition of structure to a statistically averaged approximation. In the case of polar liquids the situation is further obscured as molecular interactions are a directed process governed in large part by the molecular electric field [2,3]. In the case of liquid water the high mobility of the hydrogen nucleus additionally compounds the problem as nuclear quantum effects become non-negligible[4–6] and one must be careful in assigning simplifying assumptions, especially in the case of non-equilibrium systems [7,8]. The microscopic structure of water has been a source of continuous scientific debate and even today continues to challenge not only our instrumental capabilities [9] but also our theoretical framework within which we interpret the data [10–12]. Despite decades of effort the question remains as to what meaning, if any, can be derived from established structural elucidation techniques like neutron and X-ray scattering in part because of the degree of variability reported in the literature and even between identical measurements [13–15] begs the question whether or not a new approach is necessary [16,17].

The situation has been helped by advances in computer simulation which permit the determination of local static structure using computationally lighter pair-wise interaction potentials. Recent efforts have focused on the use of explicit three body interaction potentials [18,19] in order to more accurately predict dynamic physical quantities such as dielectric permittivity [20] and to recover infrared spectral features [21,22]. The local molecular geometry and interaction dynamics are mutually deterministic and for polar molecules like water establishes the potential energy surface upon which hydrogen bonding occurs[23]. Developing a robust model can reveal the mechanism by which short range correlation gives rise to long range cooperative order; information which is vital to advancing biology and related fields as solvent interactions play a critical role in shaping and running biochemical machinery [24–26]. Electric field intensities on the order of tens of mV/nm have been measured in living cells [27], and are common in the environment (e.g. around clay soil particles [28], solvated ions [29], and atmospheric weather [30,31]).

An essential but more poorly understood phenomenon is the effect that moderate electric fields (mV/nm) have on the local structure of the hydrogen bond network in liquid water [32–34]. Dielectric relaxation (DR) is dependent on the structure and composition of the material being subject to an electric field as well as the field intensity and frequency [35]. For simple apolar dielectric materials the relaxation response is solely dependent on the decay kinetics of induced dipoles [36]. However, in the case of polar liquids such as methanol, DMSO, and water the situation is more complex. The interaction of the permanent molecular dipole, induced dipole moment, and hydrogen bond mediated coupling must be considered [22]. Rather than being attributable to a purely non-local process as previously thought [37,38], the DR of water exhibits a departure from Debye theory attributable to the local structure of water [39,40].

Techniques such as THz spectroscopy of water-lipid micelles [41,42]or aqueous fluorescence spectroscopy [43] can probe the microscopic DR of water, however, the addition of molecular probes or confinement schemes introduce additional effects such as hydrophobic interactions. Direct examination of the effect an electric field has on local structure can improve our understanding of the link between the microscopic and continuum level DR of polar liquids such as water.

The floating water bridge is an intriguing phenomenon that occurs when a high potential difference (kV $cm^{-1}$) is applied between two beakers of water[44]. Induced by the field, the water jumps to the edges of the beakers and creates a free hanging water string through air connecting the two beakers. An example of a heavy water bridge is shown in figure 1. Despite of its ease of generation, the physical mechanism behind the formation of the water bridge and its relation to the microscopic properties of water are not completely understood. More generally water bridges belong to the broader class of electrohydrodynamic (EHD) liquid bridges which can be formed from a large number of polar liquids[45,46]. They are also characterized by a number of interesting static and dynamic phenomena [44,46–50] including optically active transient structures which result from perturbation of the local refractive index and hint at changes in local dipole magnitude[51].

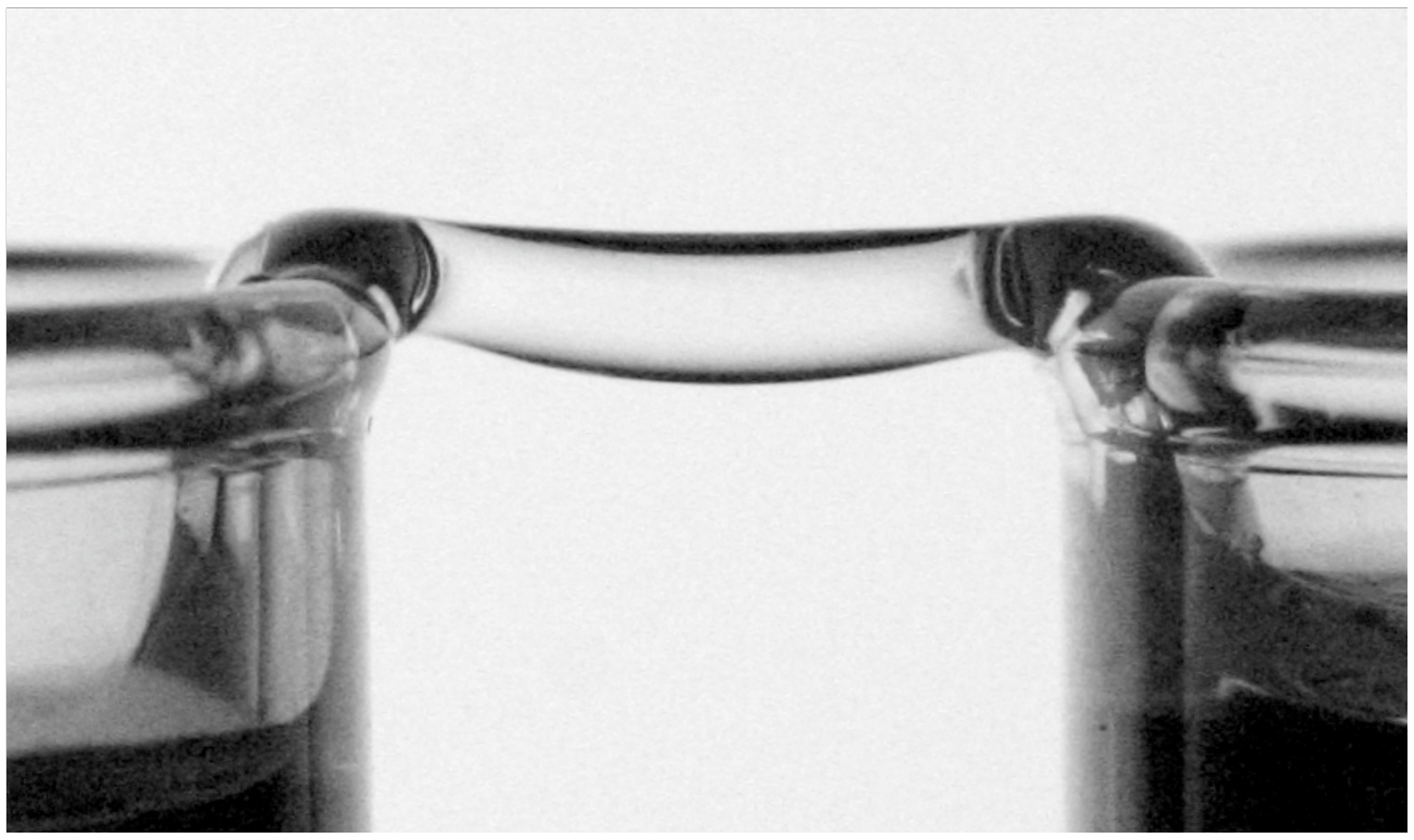

*Figure 1: $D_2O$ bridge (l ~ 10mm, d ~ 1.8 mm) using deionized D2O (* $\square < 1\mu S / cm$), 15kVDC

EHD bridges provide experimental access to a unique non-equilibrium liquid state [52] and offer the opportunity to examine how structure and dynamics in liquids are affected by a moderate electric fields. These bridges operate in the same mV/nm range as natural systems of interest and are well below field strengths which can trigger a phase change [53] and can be used to extend the phase diagram with a third coordinate[54]. Previous neutron scattering experiments [55] showed no difference in the microdensity of a $D_2O$ bridge compared to the bulk and anisotropic neutron scattering experiments [56] indicated a preferred molecular orientation within a floating heavy water bridge; however a later small angle X-ray experiment found no preferred molecular orientation and showed that measured differences were due to temperature [57]. Experimentally, meso- and microscale structure factors are accessible by neutron and X-ray scattering measurement techniques which are easily applicable to this experiment since the water cylinder forming between the two glass beakers floats freely in air. In this work the static structure of the liquid within EHD water bridges derived from combined x-ray and neutron scattering with a focus on the molecular scale ($q$ > 0.5 $Å^{-1}$) is presented.

**Methods**

*Small angle neutron scattering*

The neutron scattering experiments were carried out at the ISIS pulsed neutron source, Rutherford Appleton Laboratory, United Kingdom, on the neutron diffractometer SANDALS allowing the observation of large scattering vectors with a high resolution. The wavelength of the incident neutrons was between λ=0.05 Å and λ=4.95 Å, the dimensions of the beam were 7 mm horizontal x 20 mm vertical and the scattered intensity was measured with a ZnS detector which has 660 cells. Due to the time-of-flight principle, momentum transfers from $q$=0.5 $Å^{-1}$ to 50 $Å^{-1}$ are accessible. The floating aqueous bridges with 10 mm in length were set up as described in [58] with glass beakers using a specially constructed remote-control and monitoring set-up inside the SANDALS sample chamber. This allowed reproducing the beaker position and the bridge length with a precision of < 0.1 mm. A sketch of the measurement geometry is given in figure 2.

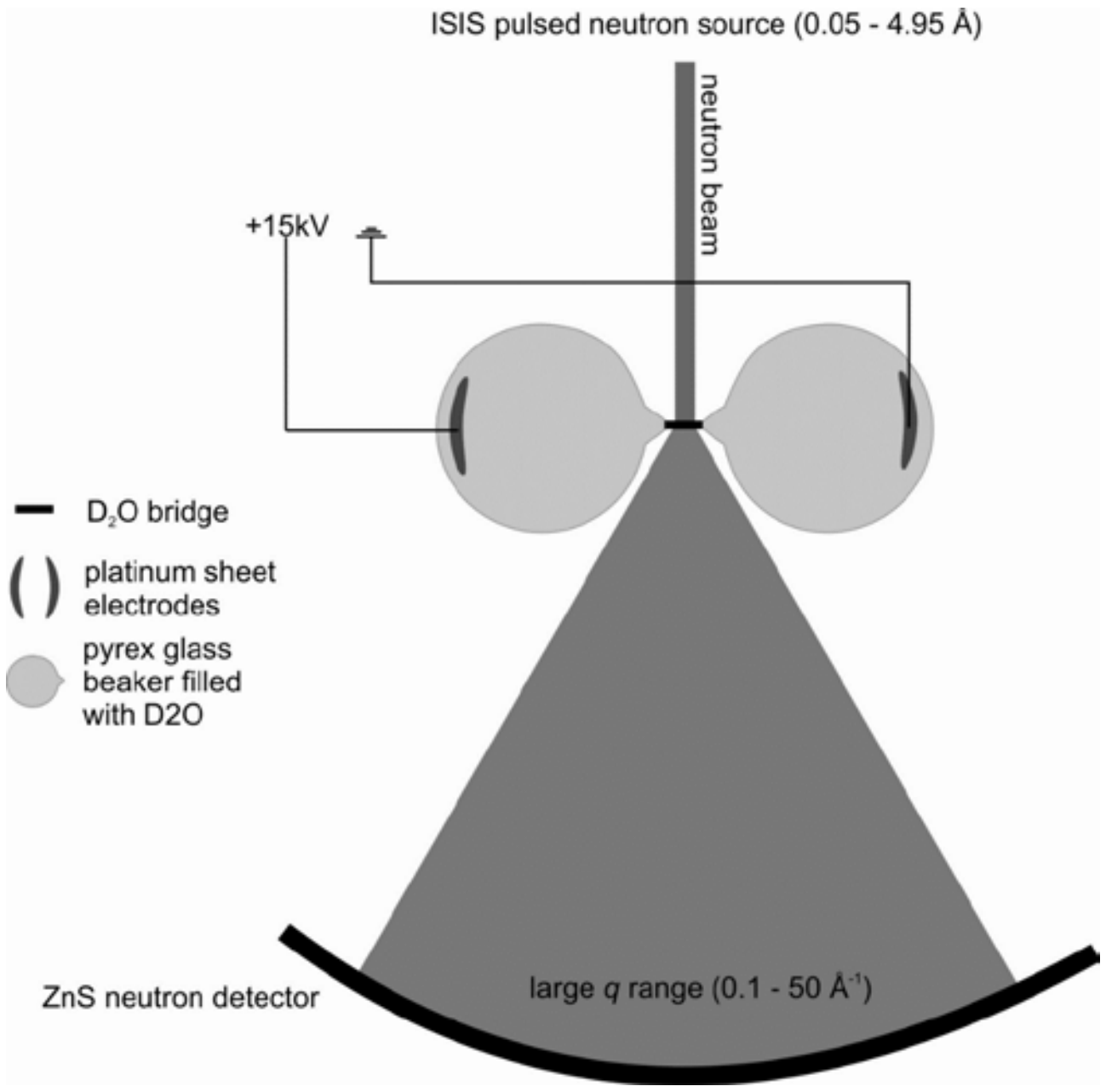

*Figure 2: The aqueous bridge set-up at the SANDALS instrument.*

The bridge system during operation is considered as three connected liquid bodies: the anolyte, the catholyte, and the bridge itself. After bridge operation ceases only the former two bodies remain. Electrolysis, though retarded by the low ionic conductivity, does result in 0.5 pH units difference between the anolyte and catholyte [59] and it was necessary to characterize how this imbalance could affect the measured neutron scattering profile.

The water bridge formed between two beakers placed horizontally on either side of the neutron beam as soon as the voltage was applied. Operating at ~12-15 kV, the bridges that formed looked alike for light water ($H_2O$), heavy water ($D_2O$) and a "null" mixture (0.6407 $H_2O$ : 0.3593 $D_2O$) of light and heavy waters with diameters oscillating between 1 and 3 mm. The use of isotope substitution in neutron scattering improves the partial structure factor resolution. In null water, the hydrogen atom has zero scattering length on average, thus in principle the O-O correlation can be measured directly. However, the O-O signal constitutes only 12% of the total signal compared to the incoherent background from the hydrogen, thus a very high signal-to-noise ratio is required for this measurement.

The experiment was run in air, since an evacuation of the chamber was, for obvious reasons, not possible, and previous studies [60] have shown that the bridge is unstable or even unachievable in noble gas atmospheres. Thus, a large volume of air (~400 mm diameter) had to be traversed by the neutron beam, creating a large background scattering which significantly reduced the quality of the data from the water bridge.

A horizontal vanadium rod (6 mm diameter) was used to calibrate data on an absolute scale. The beam size was set to 7 mm horizontal x 30mm vertical, using adjustable slits in the beam line and a fixed aperture which was inserted just before the sample. In order to verify the calibration procedure a rod of amorphous $SiO_2$ with 4 mm diameter was placed between the beakers at the exact position of the water bridge. Figure 3 shows the structure factor from this rod, after corrections, and compares with data obtained from a 3mm solid slab of silica measured on SANDALS under the same conditions as the present study, but using the full beam size, 30 x 30 mm. This measurement confirms that accurate data are obtainable using the water bridge geometry, but also shows that the statistics are likely to be poor due to the reduced sample volume in the beam (about 0.1 mL in this case) and the large air scattering from around the bridge. For the water bridges, the diameter was typically smaller than that of the silica, so that the volume of the sample in the beam was even less, around 0.03 mL, which is about 40 times smaller than is normally used on SANDALS.

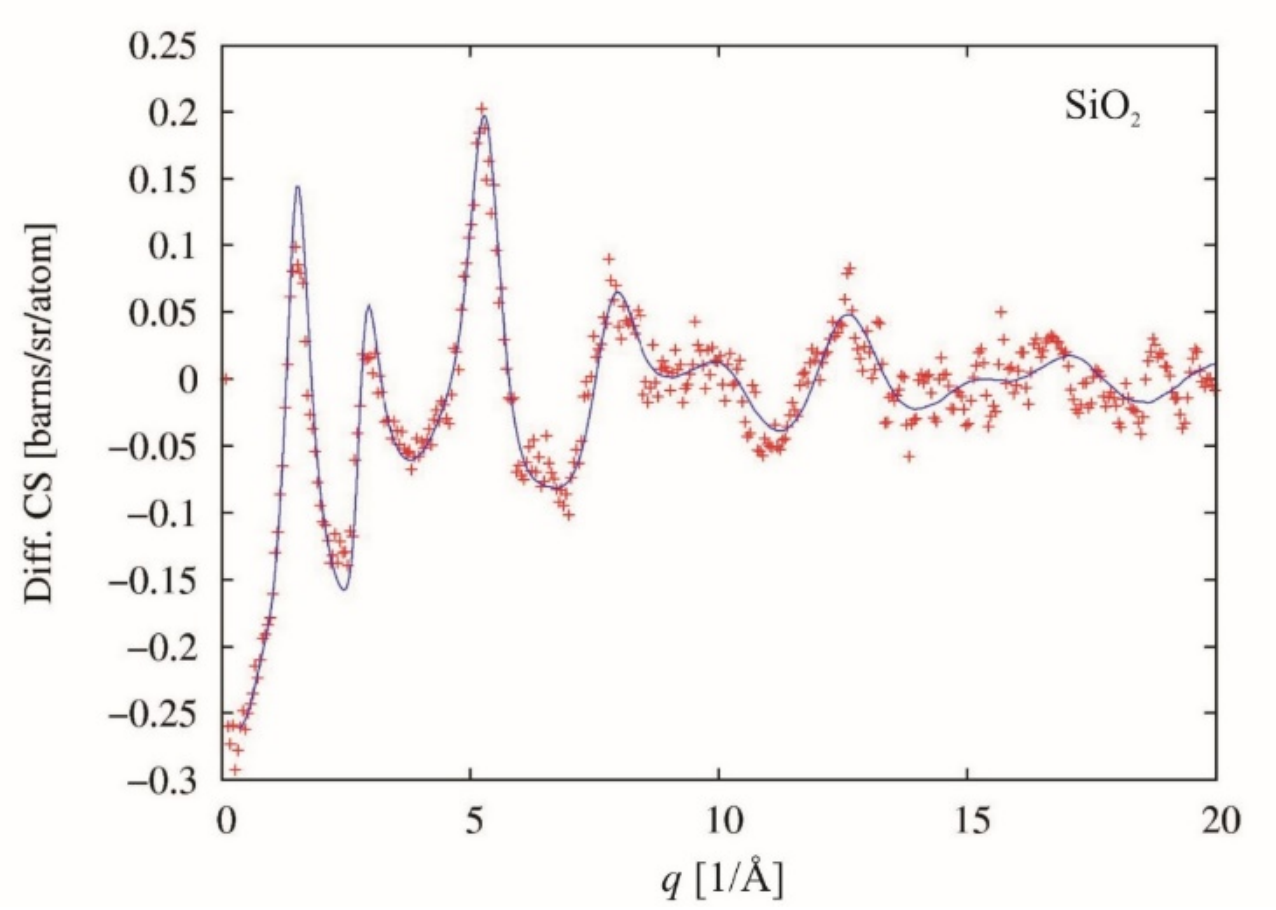

*Figure 3: Interference differential scattering cross section for a horizontal amorphous silica rod 4mm in diameter placed at the same position as the water bridge (red crosses). The corresponding data from a 3mm slab of amorphous silica is shown as the solid blue line.*

The corrections for normalizing to the incident beam monitor, background subtraction, calibration against the vanadium sample scattering, and corrections for multiple scattering and attenuation were applied using the GudrunN [61] software package. The data were analyzed detector by detector, and only added together at the very end. About 5% of the detectors showed low counts due to electronic faults or masking by the vacuum tank and were therefore excluded from the analysis.

Due to the dynamic nature of the phenomenon, the diameter of the floating bridges changed from 1 to 3 mm. Thus, in order to put the diffraction data on an absolute scale, it was necessary to assume the density of the bridge water was the same as bulk water, and use the diameter as an adjustable parameter to give the correct overall scattering level. At this point it should be noted that a previous neutron scattering study showed that the micro=-density of a $D_2O$ bridge is sensibly the same as the bulk $D_2O$ density [55]. Therefore using bulk density for this calculation is a valid assumption.

The average diameters obtained over several bridges and many (e.g. 8+) hours of up-time were 1.60 mm ($H_2O$), 1.80 mm ($D_2O$) and 1.56 mm (null water). These diameters gave scattering levels typical to those found in similar experiments on the bulk versions of these samples. Typically, the scattering level for light and null water is a few % below the theoretical limit due to inelasticity effects at the larger scattering angles.

*Anolyte and Catholyte Neutron Scattering Cross-section*

In addition to the measurements on the water bridge itself, samples of water were taken from the beakers after running the bridge and were subject to a standard diffraction measurement on the new NIMROD diffractometer in order to check whether the observed differences [59] in the anolyte and catholyte pH before and after bridge operation affected the scattering profile; this comparison is shown in figure 4.

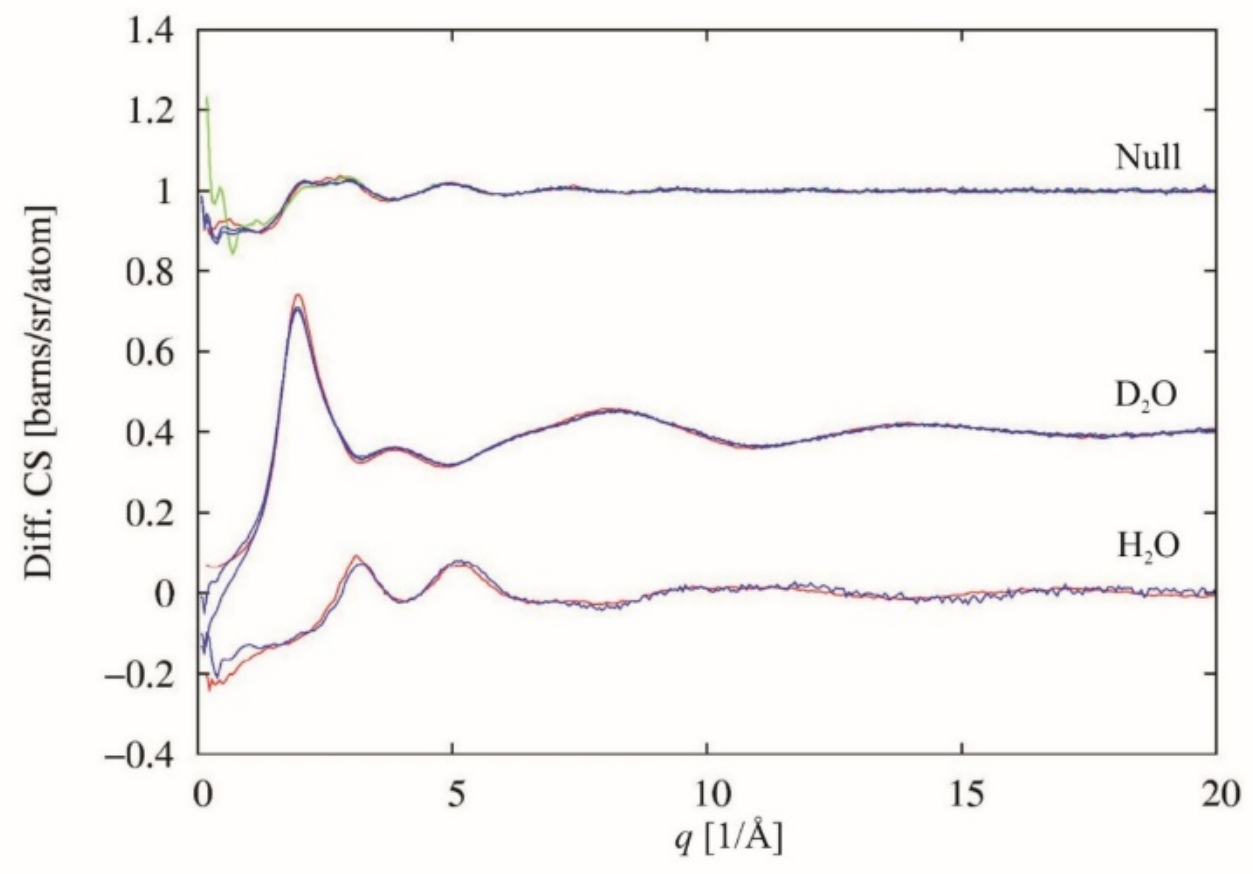

*Figure 4: Diffraction pattern from the water in the beakers after the bridge operation. The spectra from the grounded beaker and that from the HV beaker are shown as solid blue lines as measured on NIMROD, while earlier results from bulk water on SANDALS ($H_2O$, $D_2O$ and null) are shown as the red lines. For $H_2O$ only the catholyte data is shown, for $D_2O$, both electrolyte solutions are plotted, and for null water, the green line represents a bulk water spectrum run on NIMROD.*

The small differences between the anolyte, catholyte, and bulk water are due to differences between SANDALS and NIMROD instrument performance and not due to structural changes in the water. The structure below ~1 $Å^{-1}$ should be ignored as it is due to mismatch between different detector banks. This mismatch is particularly true for the NIMROD data since the instrument was being commissioned at the time the measurements were being made so the low *q* region was not fully optimized.

For these measurements the samples were contained in closed 1 mm flat plate cans, so the thickness was known within a few percent. Since the scattering level for $D_2O$ and null water was exactly as expected for these samples, there cannot have been any significant exchange of $H_2O$ for $D_2O$ (e.g. light water contamination) while the bridge was running.

The discrepancies between the different measurements are within the experimental uncertainties. When, e.g., a certain pressure is applied or solutes added [62,63] to water, the trends in the structure factors are clearly visible which is not the case for the data presented. Therefore, we conclude that the water in the beakers after bridge operation is, with regards to this measurement and within the detection limits, identical to bulk water. It is noteworthy that there was no change in the overall scattering level of the null water from the two beakers, the scattering levels are equal within 1%, indicating there was no discernible electrophoretic separation of $H_2O$ from $D_2O$.

Discrepancies between the SANDALS data of 2007 and the HASYlab refinement can be attributed to the different starting potentials used in the EPSR simulation since the HASYlab refinement uses opposite charges on the H and O atoms whereas the present refinements use no charges[64].

*X-ray powder diffraction*

X-ray powder diffraction measurements were conducted on the MS-X04SA beamline[65] at Paul-Scherrer Institute (Vilingen PSI, Switzerland) using the 60000 element MYTHEN II strip detector [66] which collects up to 120° of scatter simultaneously. Radial translation of the detector removes small gaps of 0.17° between modules and acts to improve intensity normalization as different channels will collect the same scattering angle. For the measurements used here the detector assembly was translated to four radial positions allowing collection of scattering angles between 0.105° and 125.161° ( $10^{-3}$ Å$^{-1}$ > $q$ < 22.5 Å$^{-1}$). The monochromator was set to a beam energy of 24.9882 keV (λ=0.496Å). The incident beam was focused into the center of the bridge and slitted down to cover a rectangular area 1.00 mm vertical by 4.00 mm horizontal. Type 1 ultrapure water (Cat No. ZMQSP0D01, Millipore Corp., MA, USA) was used throughout the X-ray experiments for both bridge and reference measurements. The bridge experiment was set-up in a similar fashion to that of the neutron experiments and is shown in Figure 5.

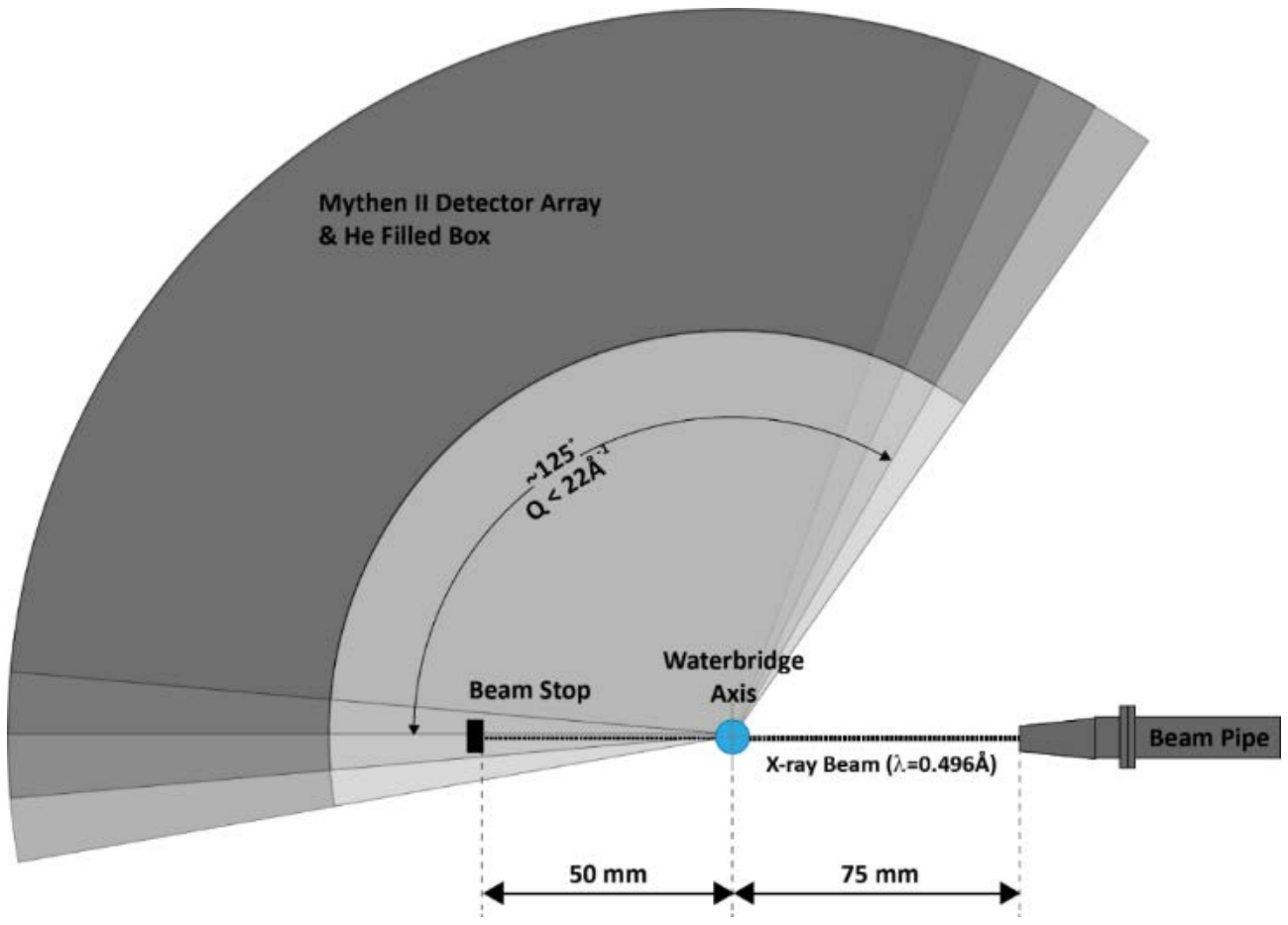

*Figure 5: Schematic representation of the experimental setup. The beakers and electrodes are not shown for clarity.*

The beakers were mounted onto an armature with two moveable platforms (Cat No. MTS25Z8, Thorlabs, USA) that allowed both bridge extension as well as centering the incident beam so it coincided with the center of the bridge. The armature was mounted to the two axis mechanical translation stage integrated into the beamline endstation; this allowed vertical translation of the bridge and alignment with the beam focal point. Platinum sheet electrodes (Cat. No. 7440-06-4, Alfa Aesar GmbH, Germany) delivered the positive 9-18 kV generated by the high voltage power supply (HCP 140-20000, FUG Elektronik GmbH, Germany). Typical current consumption was 400-1200 µA for bridge diameters between 3-8 mm and lengths between 6-8mm. Bridges are known to undergo an initial stabilization period of approximately 15-20 minutes during which flow and thermal variations are maximal [67]. Measurements commenced 30 minutes after ignition once these inherent instabilities had subsided and continued for as long as the bridge remained stable – determined to be the period where the bridge diameter reduction remained less than 20% of the initial value. Reference scattering at 0V was collected from a cylindrical jet of water gravity fed from a 20L reservoir maintained at a range of constant (±1.5°) temperatures between 5°C and 65°C. The flow rate was regulated so the jet overfilled the beam aperture in the same fashion as the water bridges.

The differential scattering cross-section was calculated from merged normalized data sets for the measured total scattering profile using GudrunX [61]. The background scattering correction was derived from the empty beam (air) with the armature in place. The Fourier transform over the measured *q* range gave reliable results of the radial distribution function between 0.3 and 10 Å.

*Temperature effect on total X-ray scattering and tetrahedrality*

Temperature affects the density of liquid water and thus the effect on the scattering profile is a non-trivial manner[6869]. The first two peaks from the corrected X-ray scattering data of the reference jets and the combined water bridge samples were fitted as a function of temperature (Fig. 6). Temperature references were prepared to compare the bridge to. However, it is known that the actually temperature in an operating bridge will fluctuate by as much as 10 degrees and that local heating may transiently occur. Thus rather than attempting to fit the temperatures directly to the position of the first two S(Q) peaks, $S_1$ and $S_2$, as has been done previously [28] a different approach which checks that the relationship between the magnitude of the second g(r) peak, $g_2$, as a function of the splitting, Δq, between $S_1$ and $S_2$.

The collected data is in agreement with a previous small angle X-ray scattering study [57] and shows that for the extracted scattering data the principle difference in position and intensity correlates well with temperature effects. This correlation reveals that the effect of temperature on physical quantities such as density are within the expected bounds and helps in providing a good point of comparison with reference scattering intensities and subsequent derived quantities from water samples under zero applied potential.

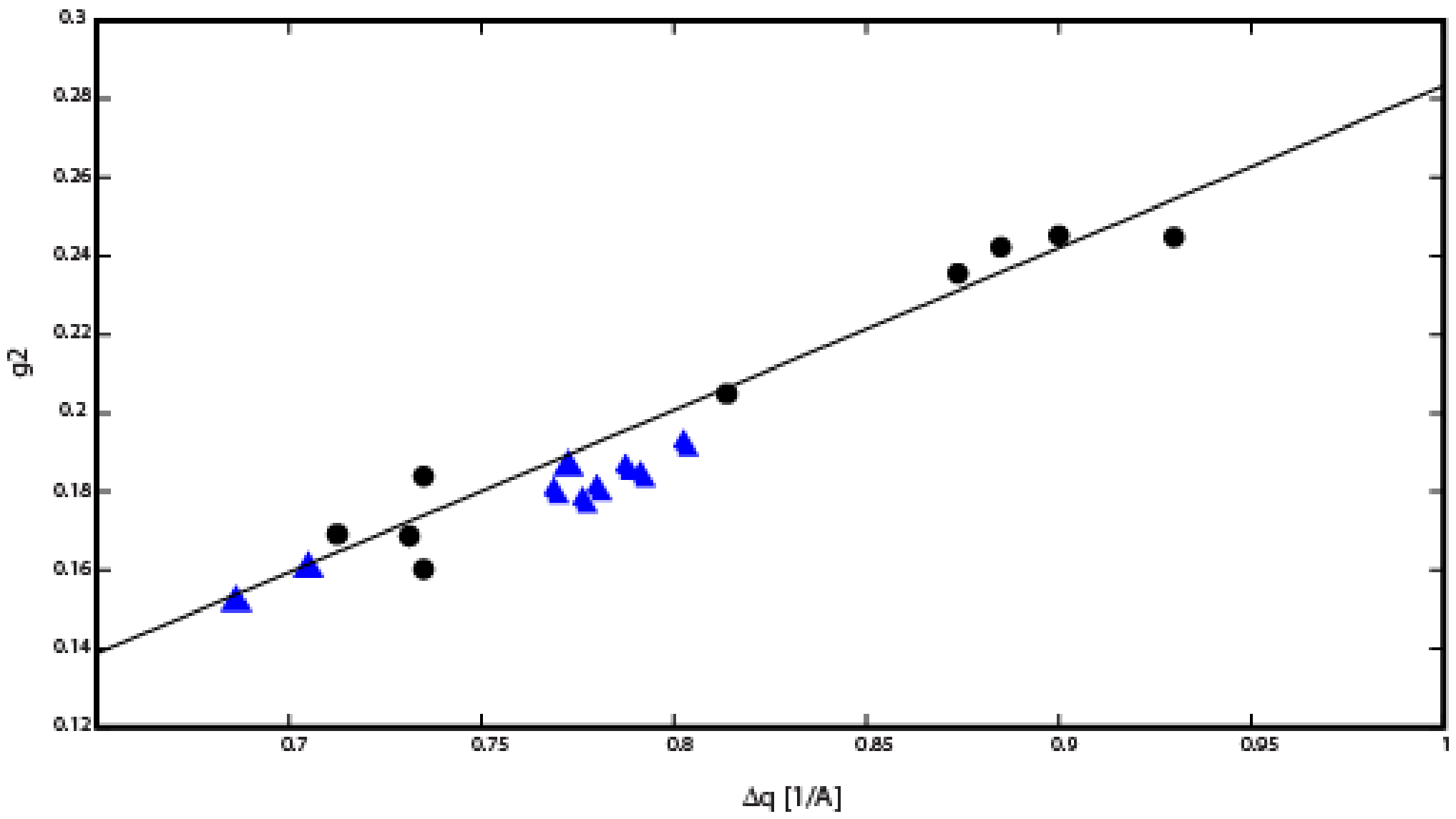

*Figure 6: A check of whether the temperature dependence on tetrahedral order holds in the presence of moderate intensity electric fields. Water bridge data (triangles) is in good agreement with reference data (circles). The line is a guide to the eye and fit from only the reference data ($R^2$=0.9375).*

*Empirical potential structure refinement*

The method and limitations of analyzing combined X-ray and neutron data from disordered materials is covered extensively in the literature [6,7,70]. The methods used herein are consistent with those used for liquid materials and specifically water. In the case of the use of an EHD bridge as sample it bears to briefly remind the reader of potential pitfalls in using the standard approach. The detector geometries are somewhat biased in relation to the imposed electric field parallel to the longitudinal bridge axis. On average the neutron scattering vector is along the bridge axis, whereas X-rays scatter in a perpendicular direction. However, the neutron detectors on SANDALS are designed for large square samples and thus span a range of azimuthal angles, depending on the scattering angle, which means that in practice there can be quite a few off-axis scattering vectors included in the average signal. This problem is essentially non-existent in the MYTHEN II detector as this system is optimized for cylindrical samples and covers an azimuthal distance (8mm) which is close to the incident beam width (4mm). The atomic form factors, inelastic scattering contributions, and material density are considered unchanged from that of bulk water throughout the data reduction and subsequent analysis. Empirical structure refinement was conducted using the EPSR24 distribution [71] using the TIP4P/2005 force field [72]. This model has been previously shown to quantitatively agree with small angle X-ray scattering data [73], matches well to key physical parameters such as the density maximum of water [72], and has been used for other combined neutron X-ray refinement studies [64,74] and explorations of intermediate-range order in water [75]. The simulation box contained 1000 molecules with a number density of 0.1002 molecules/Å$^3$. The molecules were randomized and brought to equilibrium without the application of the empirical potential for 100 cycles. Identical copies of this box were loaded as the starting point for the simulations were an empirical potential was applied to force the scattering profile of the simulation to approach that of the measured differential scattering cross-section. Both simulations ran simultaneously on the same computer so that the effect of the pseudo random number generator producing artifacts would be minimized. The simulations were also run for a large number of accumulations to reduce the output to a statistically significant and stable configuration. For every five steps in the simulation an output file was recorded and these were compounded into the stable configuration which included 23650 snapshots.

**Results and Discussion**

*1. Total scattering of water under moderate electric field*

A comparison of the corrected differential scattering intensity from both neutron and X-ray experiments for bridge and bulk water is shown in figure 7. Differences are evident in the scattering signal in the range 1-5 $Å^{-1}$. Small changes in intensity for the X-ray data are evident for larger *q* range whereas at low *q* the departure between the two data sets becomes more pronounced. The first peak in the heavy water neutron scattering shifts to higher *q* in contradiction to previous work [55]. Oscillations at high *q* are also more pronounced in the case of the electrically stressed bridge water. The electric field steepens the transition between the first and second peaks in the neutron scattering data from light water. Null water in the bridge is particularly interesting since it shows very little of the characteristic oscillatory structure up to 10$Å^{-1}$ which is present in the bulk liquid (reference water, no electric field). This is usually indicative of the pronounced O-O distance of ~2.8Å which is seemingly less pronounced in the electrically stressed bridge water. Another feature is that for both $H_2O$ and $D_2O$ the high *q* oscillations do not decay quite as rapidly with increasing *q* in the bridge compared to bulk. The Fourier transform of the scattering intensity provides the radial distribution function (RDF) (figure 8). The previous features are again visible, namely a slightly weaker intensity in the first X-ray peak along with reduced oscillations between 1-3 $Å^{-1}$. The first peaks for both heavy and light water are both stronger in the bridge, and again the oscillations up to 3 $Å^{-1}$ are more strongly pronounced. These effects are known but not satisfactorily understood within the field and are often attributed to problems in the inelastic correction. A novel feature is visible in the null water transform near 2.2 $Å^{-1}$, a feature which is also more enhanced in the light water scatter from the bridge.

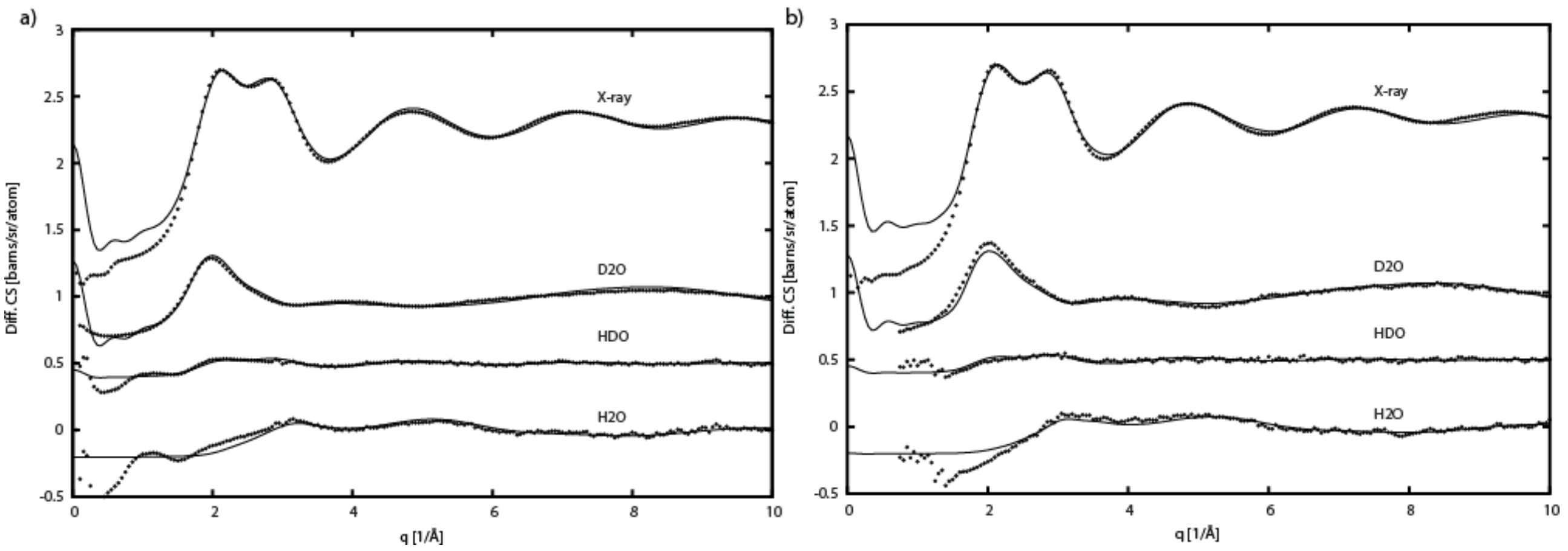

*Figure 7: Comparison of the corrected scattering data (open-circles) and the EPSR generated fits (lines) for a) ground potential liquid water at 40°C and b) electrically stressed bridge water at 40°C. The neutron scattering data from (a) was collected on the NIMROD instrument using a can geometry, while that of (b) on SANDALS with the cylindrical bridge geometry. For both cases the X-ray instrument was identical (MS04) as was the sample geometry (e.g. cylindrical bridge or jet). Close inspection of these curves reveals some misfits, but these fits appear to be the best that can be achieved with the present programs and data analysis.*

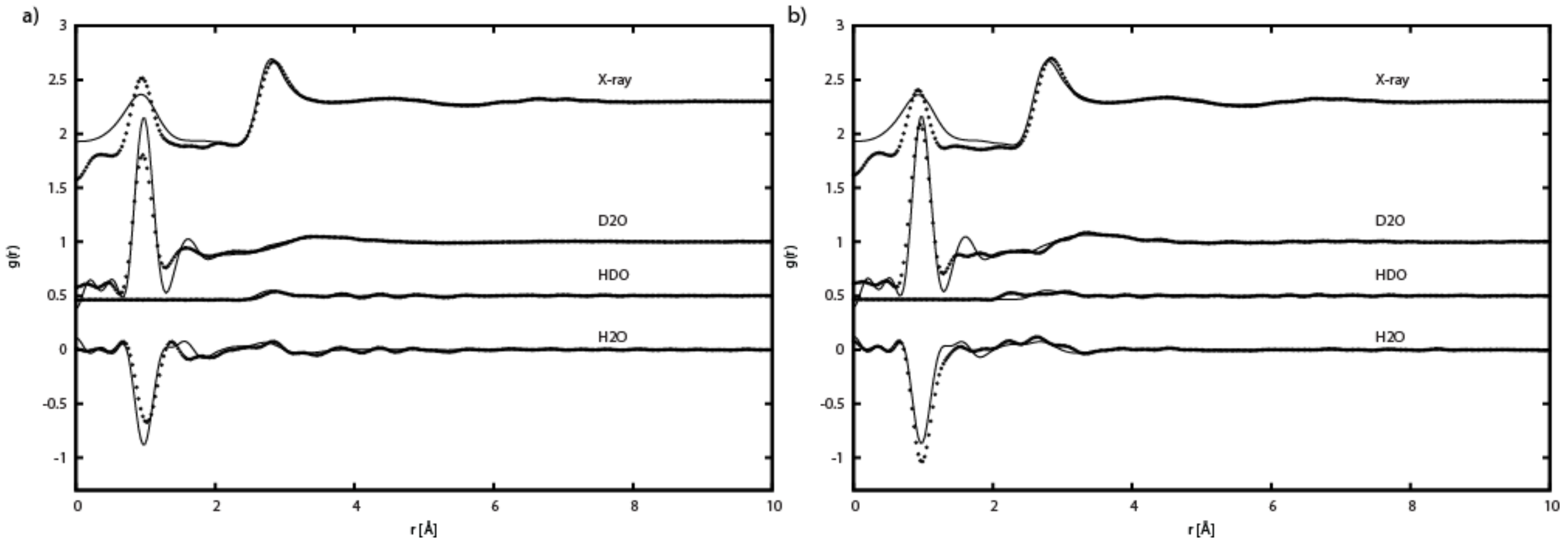

*Figure 8a: Radial distribution functions derived from the Fourier transforms of either the EPSR generated fits (lines) or scattering data (open-circles) from a) ground potential water (can or jet) at 40°C for X-ray and neutron (NIMROD) or b) from electrically stressed bridge water at 40°C for X-ray and neutron*

*(SANDALS). The small misfits carry over from the scattering data and again appear to be the best that can be achieved with the present programs and data analysis. The RMS noise in g(r) is 1.3E-04 for reference water (a) and 1.2E-03 for bridge water (b).*

*2. Structure refinement to extract partial structure factors*

The combined X-ray and neutron data sets discussed previously form a complete basis for the determination of the order parameters within liquid systems [64] and thus it is possible to extract the site-site partial structure factors, namely H-H, O-H, and O-O for water. Comparison of the simulated and measured scattering data shows a reasonable level of agreement for both the water bridge (Fig. 7b, 8b) and bulk water samples (no electric field, Fig. 7a, 8a). Again, the small discrepancies are likely due to poor inelasticity correction. Furthermore, the agreement for the respective RDFs is acceptable. The differences between fit and data for the first peaks can, in part, be attributed to the structural differences expected between heavy and light water as in the simulation they are assumed to be structurally identical. The variance in oscillatory behavior is likewise partially attributed to the degree of disordering between isotope variants; where light water is expected to be the slightly more disordered species [76]. The enhanced oscillations at high *q* may also indicate the onset of long-range dynamic reorganization in advance of electrically induced phase transitions [34,77] indicative of increasing local structure which generates medium range order even in the face of elevated temperature [75].

The relative differences between scattering with and without electric field can be discerned in the partial structure factor (PSF) distributions shown in figure 9. One reason for examining the PSF is that these distributions, unlike the RDF and differential scattering intensity, are largely unaffected by changes in temperature [62,78]. The O-O PSF for the electrically stressed bridge water is shifted to shorter distances, a trend observed for all three PSFs. The second peak in the O-H PSF also has increased intensity and a narrower distribution a feature shared by both of the first two peaks in the H-H PSF. Some of these shifts are reminiscent of liquid water at elevated temperatures and pressures well above the liquid critical point where the PSFs are compressed in reflection of the partial collapse of the molecular coordination [79] but the bridge data does not show the peak broadening also associated with pressure effects.

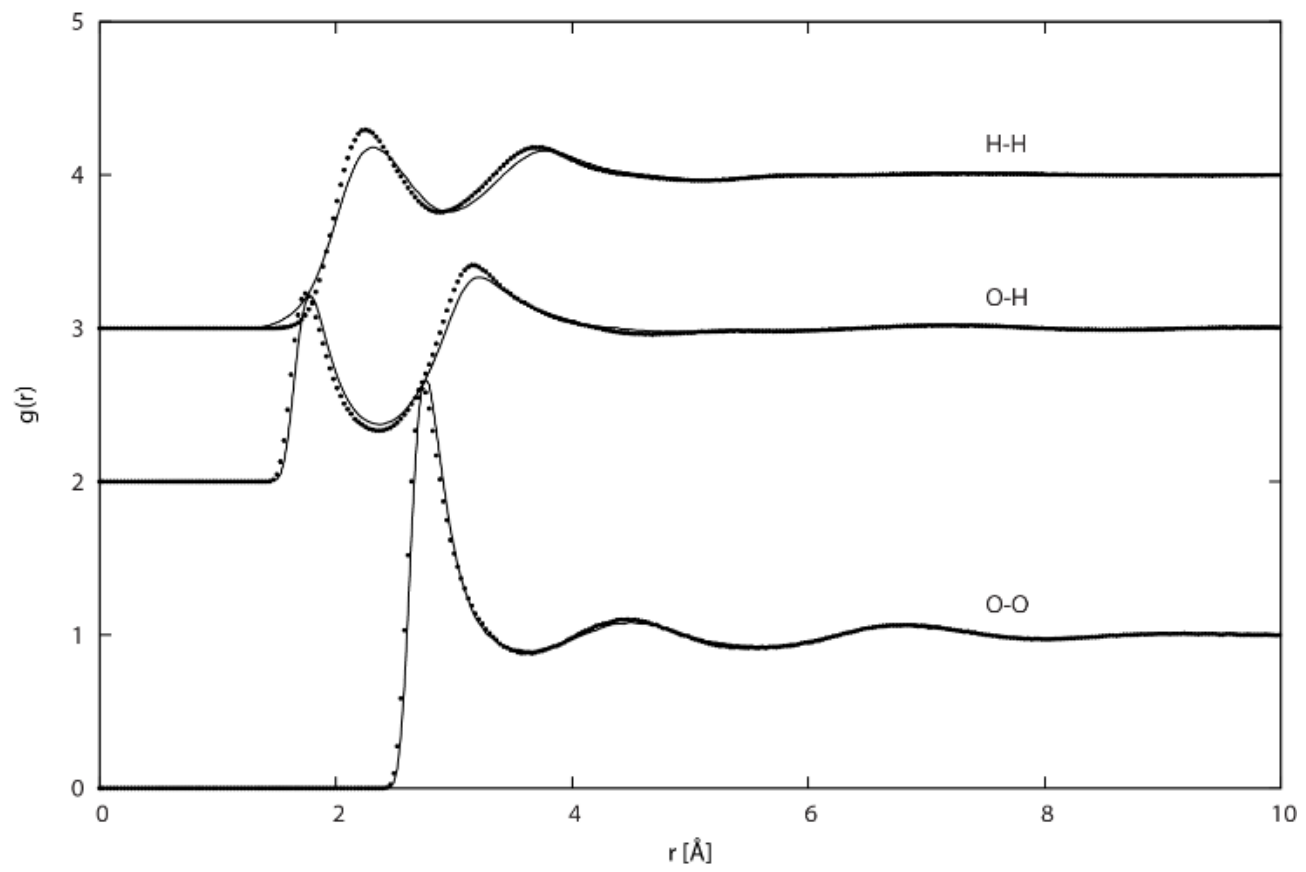

*Figure 9: Comparison of the PSFs recovered from the EPSR simulations for electrically stressed bridge water data (symbols), bulk reference without electric field (line). Plots are offset for clarity. While differences are readily observable the shifts are within the recognized uncertainty of the applied methods.*

The measured PSFs also show similarities to those reported for concentrated hydrogen chloride solutions [80], but it is uncertain whether the proton concentration in the bridge is high enough [59] to change the local structure. Solvated protons carry an intense electric field around them which distorts the local intrinsic electric field. It has been suggested that solvated protons act as an inverse-polaron [81,82] whereby the proton defect perturbs the local order and it is possible that the applied electric field is behaving in a similar way – reducing the interaction distance between molecules by strengthening the effective dipole via the action of induced dipole moments in the neighboring molecules as described by Lorentz-Lorenz theory and discussed for water ice $I_h$ [83]. The resulting local anisotropic distribution of the electric field polarizability will distort the local dipoles establishing strongly coupled local structures. Changes in the local environment will in turn distort the intramolecular potentials which cannot be captured by a rigid water model but are nonetheless captured in the raw scattering data as small changes in high *q* oscillations (see figures 7a and 7b). The implications of changes in molecular geometry by a moderate electric field are indeed surprising and illustrates that the assumptions under Born-Oppenheimer provide an incomplete basis for systems like water. The imposed electric field, though weak compared to the local interaction potential, will polarize a subpopulation of dipoles leading to further distortion of the local electron density. This will have a non-negligible effect on the light hydrogen nucleus and thus the two fields can no longer be considered as decoupled. The recovered HH pair potentials hint at this though not significantly, however it must be kept in mind that the PSFs are modeled parameters and differences in experimental data must be considered alongside the simulation results.

The PSFs are the convolution of two quantities: the imposed reference potential (*RP)* and the empirical potential (*EP)*. Examination of these contribution gives the opportunity to check how the simulation applies interaction forces in order to converge simulation and scattering data. Since a rigid non-polarizable model is used the extracted *RPs* are identical for both simulations and the Morse and Leonard-Jones potential are thus identical. The site-site empirical potentials ($EP_{\alpha\beta}$) shown in figure 10, however, are quite different even though the requested magnitude of the *EP* is identical for both systems. In EPSR there are no rules governing the shape of the EP and it is allowed to freely evolve to arrive at the best fit of simulation and experiment possible. Thus an examination of the empirical potential provides clues about how the energetic environment in the two simulations differ. The observed changes show three patterns: $EP_{HH}$ amplifies the second peak from the bulk (figure 10), $EP_{OH}$ is roughly anti-phase to that of the bulk, $EP_{OO}$ merges the two bulk peaks into a broad distribution. Thus, the forces acting upon the atom pairs are non-equivalent between the two simulation sets and this must be considered as deriving from the experimental data as all other parameters are equivalent. Disordered materials such as liquid water cannot be satisfactorily described using only two body correlation functions such as the PSF and work is ongoing to develop three-body correlated models for computer simulation [18].

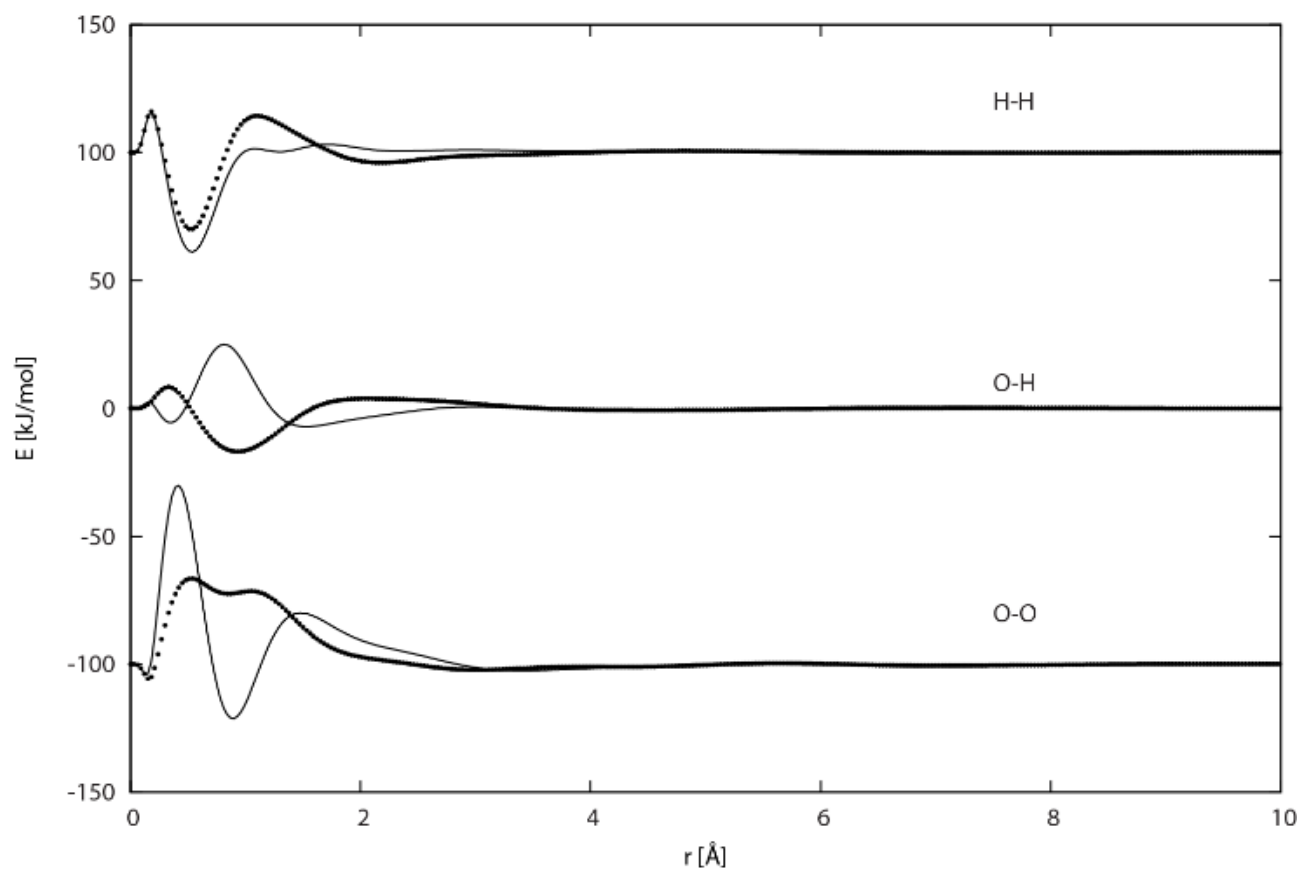

*Figure 10: Comparison of the empirical potentials from the EPSR simulations from reference water without electric field (lines) and electrically stressed bridge water (symbols). It is clear that the applied empirical potential distributions vary significantly between the two simulations. Such a mismatch is not currently reported in the literature and illustrates the underlying differences behind the smaller shifts in the PSF.*

*3. Analysis of the EPSR simulation*

The differences of the results are close to the uncertainty level, however, the simulations also provide additional information – e.g. intermolecular energy ($U_i$), pressure ($P_i$), and the chi-squared ($\chi^2$) value of the fit to scatter data – that can show non-structural differences between the two experimental cases.

These are typically monitored to check that the values obtained are realistic and the system has settled into equilibrium (i.e. not exploding nor imploding). Since the requested empirical potential is a user selected variable, the choice of which must be carefully considered, for the comparisons thus far in the work a value of $EP_{req}$ = 20kJ/mol was selected for both systems as there is precedent in the literature for liquid water and ice under varying conditions[78]. It was noticed that the systems do not settle into the same thermodynamic regime which indicated that the generated static structure changes – e.g. PSF, S(q), g(r) – were also reflected in the iterative dynamics of the simulation. It is important to clarify that the 'dynamics' discussed here are not molecular dynamics since EPSR, like the reverse Monte Carlo simulation method it is based on, lacks a meaningful time metric [84]. Rather the dynamics refer to how simulation parameters vary in response to pseudo-random perturbations. It is thus a measure of the stability of the simulation which is expected to converge if the two systems are identical. Two approaches were taken to better understand the response of the simulations to variation in $EP_{req}$: A symmetric approach where $EP_{req}$ was varied equally for both simulations between 0 – i.e. reference potential only – and 100 kJ/mol; and an asymmetric analysis of the response where the established value of $EP_{req}$ = 20 kJ/mol was held constant for the bulk water simulation and $EP_{req}$ varied between 16-37 kJ/mol for the bridge system. This range was selected as it was within a basin of stability which provided the best fit and physically reasonable parameters. The starting point for the simulations used to check the $EP_{req}$ response was an identical configuration take from the equilibrated simulation where $EP_{req}$=20 kJ/mol one for each of the two data sets. The symmetric approach was run for a minimum of 5000 iterations, whereas, the asymmetric analysis run only for 500 iterations. In both cases however, the behavior of the systems were stable within a narrow band of random oscillations as evidenced by small values of the standard error of the mean on the order of $5{\cdot}10^{-2}$ and $4{\cdot}10^{-2}$ for $U_i$ and $P_i$ respectively. As $EP_{req}$ is increased $U_i$ in both systems will be reduced as the $EP$ compensates the molecular interaction energy an action which is less effective in the bridge data as $U_i$ for any given $EP_{req}$ is always lower when no external electric field is present. In a similar manner increasing values of $EP_{req}$ drives $P_i$ ever lower indicating that the system heads toward implosion. However, in the presence of the moderate electric field the system uniquely resists this trend and $P_i$ stabilizes even at the highest requested potentials. While enticing to consider the implications of such trends in terms of the experimental system a check of $\chi^2$ also shows a quickly decaying quality of fit in the electrically stressed bridge system at $EP_{req}$ values above 30 kJ/mol. Thus, it is best to restrict our search for energetically equivalent simulations to $EP_{req}$ below this value. Here we define a new quantity – the excess empirical potential $EP_{ex}$ – which is simply the additional energy in terms of $EP_{req}$ necessary to force the two simulations to energetically converge and is defined as

$$EP_{ex} = EP_{bridge} - EP_{bulk} \, , \tag{1}$$

where $EP_{bridge}$ and $EP_{bulk}$ are the requested empirical potentials for the simulation. The response of $U_i$ and $P_i$ are different and it is thus expected that the equivalent values for these parameters may lie at different values of $EP_{req}$. Figure 11a shows the plot of $dU_i/dEP_{ex}$ yielding an equivalence point between simulations of 7-8 kJ/mol $EP_{ex}$. Figure 11b likewise shows the response of $P_i$ to $EP_{ex}$ and as expected the resulting equivalence point is around 12-13 kJ/mol $EP_{ex}$, however, the regression $R^2$ is not as good for the pressure as for the energy ~0.75 vs. ~0.94, respectively. One possible reason for the observed discrepancy may lie in the method by which these two quantities are calculated [85,86] and the assumed interactions in the TIP4P/2005 force field [87]. It may also reflect the development of dynamically heterogeneous distributions which is expected when quantum mechanical forces are turned on in the simulation[88]. These factors that will be explored in the future.

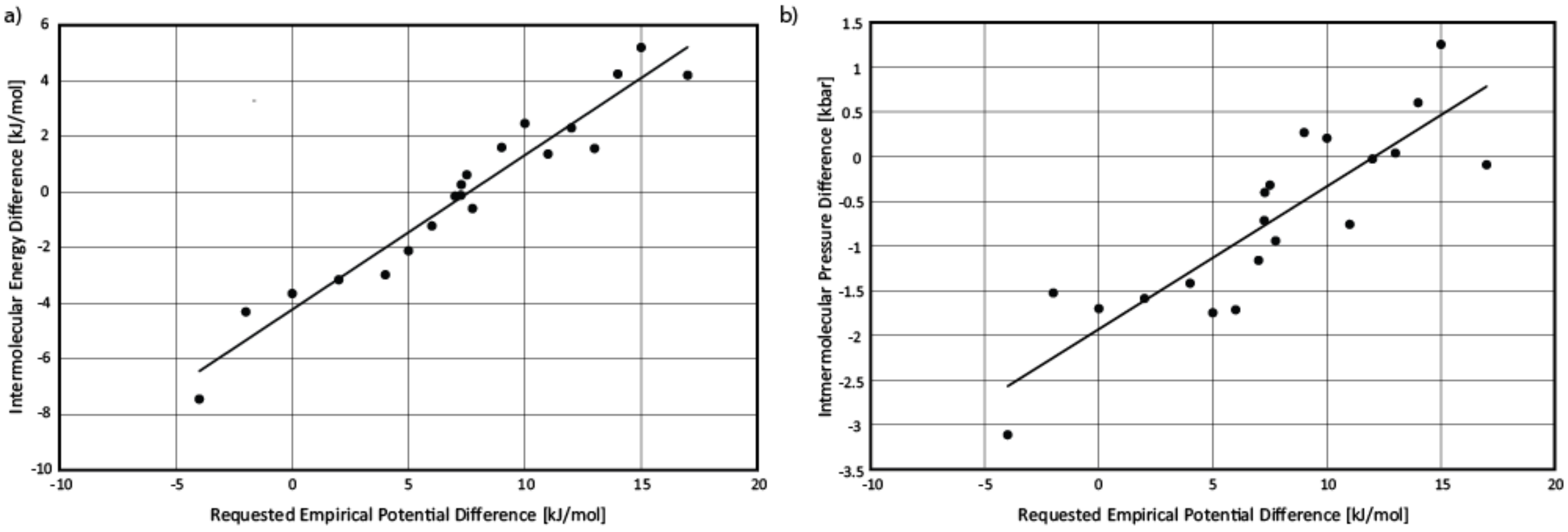

*Figure 11: A comparison of the a) differential intermolecular energy ($U_i$) or b) intermolecular pressure ($P_i$) reported in 500 configurations for difference in requested empirical potential ($d_{EP} = EP_{bridge}$-$EP_{bulk}$). An equivalency in $U_i$ is found at an excess energy of 7.5 kJ/mol requested for the bridge simulation. The pressure term requires more energy ~12 kJ/mol requested in order for the bridge simulation to approach that of the reference water. Fit values of $R^2(a) = 0.9378$ and $R^2(b) = 0.7498$ also show that the terms are not equivalent in sensitivity to noise.*

*Table 1: Comparison of the response of the simulation environment to changes in the requested empirical potential for the two simulation cases.*

| Requested Empirical Potential ($EP_{req}$) [kJ/mol] | Applied Electric Field Strength [MV/m] | Intermolecular Energy ($U_i$) [kJ/mol] | Standard Deviation Energy $\sigma_U$ | Intermolecular Pressure ($P_i$) [kbar] | Standard Deviation Pressure $\sigma_P$ | Quality of Fit $\chi^2$ | Standard Deviation Chi-squared $\sigma_\chi$ |
|---|---|---|---|---|---|---|---|
| **0** | 0E0 | -4.27E+01 | 4.30E-03 | 5.44E+00 | 8.35E-03 | 4.14E-03 | 7.91E-06 |
| **0** | 1E6 | -4.27E+01 | 4.23E-03 | 5.45E+00 | 8.01E-03 | 5.48E-03 | 9.38E-06 |
| **10** | 0E0 | -5.28E+01 | 1.65E-02 | 2.85E+00 | 1.14E-02 | 2.91E-03 | 6.87E-06 |
| **10** | 1E6 | -4.97E+01 | 1.15E-02 | 4.15E+00 | 1.09E-02 | 4.10E-03 | 7.81E-06 |
| **20** | 0E0 | -6.15E+01 | 1.98E-02 | -6.30E-01 | 1.34E-02 | 2.90E-03 | 7.00E-06 |
| **20** | 1E6 | -5.62E+01 | 1.60E-02 | 1.98E+00 | 1.40E-02 | 4.45E-03 | 9.83E-06 |
| **30** | 0E0 | -6.93E+01 | 2.56E-02 | -3.94E+00 | 1.63E-02 | 2.84E-03 | 7.28E-06 |
| **30** | 1E6 | -6.24E+01 | 1.90E-02 | -5.80E-02 | 1.53E-02 | 4.58E-03 | 9.15E-06 |
| **50** | 0E0 | -8.51E+01 | 2.68E-02 | -1.05E+01 | 1.75E-02 | 2.91E-03 | 8.20E-06 |
| **50** | 1E6 | -7.07E+01 | 1.79E-02 | 7.69E-01 | 1.70E-02 | 3.78E-02 | 1.47E-04 |
| **70** | 0E0 | -1.01E+02 | 3.30E-02 | -1.71E+01 | 1.95E-02 | 3.07E-03 | 7.72E-06 |
| **70** | 1E6 | -8.15E+01 | 3.14E-02 | -7.21E-01 | 2.06E-02 | 7.12E-02 | 4.21E-04 |
| **100** | 0E0 | -1.16E+02 | 9.07E-02 | -2.04E+01 | 6.67E-02 | 1.62E-02 | 1.30E-04 |
| **100** | 1E6 | -9.99E+01 | 2.55E-02 | -2.92E+00 | 2.15E-02 | 1.73E-01 | 4.02E-04 |

*Table 2: Analysis of the simulation outputs used for the structure refinement presented in this study.*

| $EP_{req}$ = *20 kJ/mol* | *Water bridge* | *Pure Water 40°C* |
|---|---|---|
| *Number of Configurations* | 23651 | 23654 |
| *Chi-squared* | 0.00548 | 0.00280 |
| *Repulsive pair energies [kJ/mol]* | -56.33250 | -61.34670 |
| *Harmonic pair energies [kJ/mol]* | 6.81611 | 7.22784 |
| *Number density [N/Å$^3$]* | 0.10020 | 0.10020 |
| *Noise in Average g(r)* | 0.00119 | 0.00013 |
| *R-factor $H_2O$ neutron fit* | 0.00136 | 0.00213 |
| *R-factor $D_2O$ neutron fit* | 0.00021 | 0.00081 |
| *R-factor HDO neutron fit* | 0.00031 | 0.00036 |
| *R-factor X-ray fit* | 0.02004 | 0.00790 |
| *Intermolecular pressure ($P_i$) [kbar]* | -2.60345 | -0.00235 |
| *Intermolecular energy ($U_i$) [kJ/mol]* | -90.58690 | -61.34600 |

*4. Approach to understanding total radiation scattering in electrically stressed liquid*

Determining how the structure of a polar liquid such as water will be influenced by a moderate electric field is not a straightforward task. The employed scattering methods suffer from inherent variability which is compounded by critical assumptions in data treatment and limitations of simulation based structure refinement[13]. In the experiments conducted here we must also consider the DR process which is typically considered to be a purely macroscopic phenomenon that should have no effect on local structure. However, in branching hydrogen bonded liquids such as water, DMSO, and methanol (all of which can bridge electrohydrodynamically [67]) anomalous behavior which indicates a microscopic relaxation process has been reported by studying the spectral response of small molecular fluorophores [43]. The fluctuating electric field of the relaxing fluorophore induces an extended polarization ensemble that arises from reinforced oscillations in the hydrogen bonded solvent network which differs from the bulk expectation values. The role of the hydrophobic interaction of this system is as yet unclear however DR in water under nanoscale confinement has recently been shown to be strongly anisotropic with respect to the dipole moment [89]. This supports work which showed that the departure from the Debye relaxation model arises from microscopic anisotropy in rotational diffusion [90] which is a known property of neat liquid water [91].

From the experiments performed the local structure of water appears to be unchanged and does not exhibit anisotropy within reasonable uncertainties. However, analysis of the simulation parameters and the modeled potential energy surface reveals a clear departure between the water with and without an external electric field. The intermolecular interaction potentials (Figs.10, 11 and 12) are not equivalent for a given empirical potential and furthermore the shape of the potentials are significantly different. This is not unexpected given the problems of recovering dielectric properties from simulation using rigid non-polarizable models and without the recently discussed need to correct the parameters used to define the dipole moment and polarization surfaces. Regardless of this limitation, the observed differences in the simulation behavior indicate that the applied field, although weak relative to the intramolecular coulombic field, does significantly affect the local electrical environment.

Liquid water naturally possesses a broad population distribution of molecules which are non-equivalent due to variations in the local potential energy surface (PES). The application of a moderate electric field on the order of mV/nm is insufficient to drive the liquid system into a phase change like anisotropic reorientation of the total population, however, it is sufficient to enforce local changes in the spatially heterogeneous dynamics by distorting the local spatial distribution of molecules with respect to the molecular dipole moment [92]. Unfortunately, the EPSR method using TIP4P/2005 is unable to account for polarization effects and thus the measured dipole moments do not differ between the simulations presented here. Examining the role electric fields play in local molecular distortions and the emergence of long-range order is important for resolving mesoscale dynamics in liquids.

We would like to point out, though, that the differences determined here, considering the known variability in the method as reported in the literature [13], and the observed shifts in the HH and OH PSFs, are nonetheless more pronounced in the water bridge than those reported for difference between the pure isotopes alone [93].

On the other hand it is known that TIP4P/2005 provides an incorrect description of heavy water, and while quite good at predicting the physical properties of water when compared to other models it falls short. The approach modeled here also fails to accurately reproduce the dielectric constant of water, the reason for which was recently discussed extensively by Vega [94] and which boils down to the incorrect assumption that the parameters used to define the PES, which determines the allowed steps in the EPSR simulation, can also be used to define the dipole moment surface (DMS). This mistake is pervasive throughout computational chemistry and new work is now emerging which takes the more correct approach [19,54].

From this research it also becomes evident that current methods are still insufficient to treat such non-equilibrium systems and there is a strong need to further develop simulation techniques that properly reconstruct the microscopic DR process as well as can account for an excited molecular sub-population.

## Conclusions

The total radiation scattering of water stressed by a moderately strong electric field (1mV/nm) was compared to water without a field using the standard approach to structure determination in disordered materials, namely X-ray and small angle neutron scattering. These measurements indicated that the modeled local potential energy surface reveals a departure between electrically stressed and unstressed water. The observed differences indicate that the local environment is changed by the applied electric field although being weak relative to the intermolecular coulombic field. It is discussed that the methods used to simulate the pair potentials are insufficient to treat such non-equilibrium systems and further simulation techniques have to be developed to properly reconstruct the microscopic dielectric relaxation process.

**Acknowledgements**

This work was performed in the cooperation framework of Wetsus, European Center of Excellence for Sustainable Water Technology (www.wetsus.eu). Wetsus is co-funded by the Dutch Ministry of Economic Affairs and Ministry of Infrastructure and Environment, the Province of Fryslân, and the Northern Netherlands Provinces. The authors like to thank the participants of the research theme "Applied Water Physics" for the fruitful discussions and their financial support. We also wish to thank Dennis Cowdery and Stephen M. Cox for their technical contributions to the neutron scattering experiments.